\documentstyle[prb,aps,multicol,epsf]{revtex}
\begin{document}
\tighten
\draft 
\title{
Statistics of Coulomb blockade peak spacings
for a partially open dot
}
\author{A. Kaminski and L.I. Glazman}
\address{Theoretical Physics Institute, University of
Minnesota, Minneapolis, MN 55455}
\maketitle

\begin{abstract}
  We show that randomness of the electron wave functions in a quantum
  dot contributes to the fluctuations of the positions of the
  conductance peaks. This contribution grows with the conductance of
  the junctions connecting the dot to the leads. It becomes comparable
  with the fluctuations coming from the randomness of the single
  particle spectrum in the dot while the Coulomb blockade peaks are
  still well-defined. In addition, the fluctuations of the peak
  spacings are correlated with the fluctuations of the conductance
  peak heights.
\end{abstract}

\begin{multicols}{2}

\section{Introduction} 
\label{sec:intro}

Quantum dot in the Coulomb blockade regime is conventionally described
by the constant interaction model.\cite{Kastner92} In this model, the
Hamiltonian of the system is represented by a sum of two terms. The
first one is the electrostatic charging energy, that does not
fluctuate and depends on the total number of electrons only; the
second term is the Hamiltonian of free quasiparticles (all
interactions except the charging energy are ignored). For a disordered
or chaotic quantum dot, free quasiparticles inside the dot are
described by the random matrix theory (RMT).\cite{MehtaBook} The
spectrum of quasiparticles is random, with the level spacings obeying
the Wigner-Dyson statistics. The one-particle wave functions are also
random, and their statistical properties within the RMT theory are
defined by the Porter-Thomas distribution.

The predictions of this model may be checked in transport experiments
with a dot weakly coupled by point contacts to the electron
reservoirs (leads). The number of electrons on the dot in this setup
is nearly quantized, and can be controlled by an additional electrode,
gate, which is capacitively coupled to the dot.  The dependence of the
tunnel conductance through the system, $G$, on the gate voltage $V_g$
exhibits sharp peaks. A Coulomb blockade peak corresponds to a point
where the ground state of the dot is degenerate: the states with $n$
and $n+1$ electrons have the same energy. The $(n+1)$st electron can
therefore freely tunnel to and from the dot, so $G(V_g)$ has a peak at
this point.

For an almost closed dot, the height of the peak is related to the
values of the wave functions at the points of
contacts.\cite{JalabertEtal92} The randomness of the wave functions
translates into the randomness of the peak heights.  The existing
experimental results\cite{ChangEtal96} for the peak conductance
distribution function are in a satisfactory agreement with the theory
predictions.\cite{JalabertEtal92}

In the framework of the constant interaction model, the spacings
between the conductance peaks contain two contributions.  The first
one is due to the charging energy, and does not fluctuate. The second
term in the limit of weak dot--lead tunneling is proportional to the
spacing between the discrete energy levels of the RMT Hamiltonian.
This term does fluctuate, and obeys the Wigner-Dyson statistics.
However, the experimentally measured distribution of peak spacings is
similar to Gaussian,\cite{SivanEtal96,SimmelEtal97,PatelEtal98} 
in apparent  disagreement with the predictions of the constant
interaction model. One of the possible explanations of this phenomenon
is that the electron-electron interaction, which is not accounted for
by RMT, substantially affects the dot's spectrum.\cite{WalkerEtal99}
Another explanation refers to the change in the dot shape with the
variation of the gate voltage $V_g$, so the two adjacent peaks in the
$G(V_g)$ dependence correspond to two different realizations of the
random Hamiltonian.\cite{VallejosEtal98}

The above-mentioned works concentrated on the almost closed dots, when
the tunnel junctions are used only as probes, without affecting the
states in the dot. In the recent experiment\cite{MaurerEtal99} the
statistics of peak spacings was studied for a partially open quantum
dot. In such a system, charge quantization is lifted gradually, while
the conductance of the dot-lead contacts increases.
Simultaneously, a crossover occurs between the sharp Coulomb blockade
peaks and  conductance fluctuations of relatively small amplitude in an
almost open dot. The main part of the peak spacing comes from the
charging energy. Therefore lifting of the charge quantization may
yield an additional significant contribution to the peak spacing
fluctuations. This many-body problem was not addressed theoretically
by now.

In this paper, a theory of the fluctuations of the peak positions in
the case of a partially open dot is developed. In Sections
\ref{sec:posit} and \ref{sec:statprop} we consider the regime in which
the conductances of the dot-lead point contacts are not negligible,
but still considerably smaller than the quantum unit
$G_0\equiv e^2/\pi\hbar$.  Under these conditions, the conductance
peaks remain well-defined.  However, their positions are already
affected by the finite dot-lead coupling. The randomness of the
electron wave functions controlling the coupling, contributes to the
fluctuations of the peak spacing. We calculate the variance of this
contribution, and show that it increases with the junction
conductance. We study the fluctuations of spacings between the
adjacent, as well as between distant peaks. This theory allows us to
form a consistent understanding of the experimental
results\cite{PatelEtal98,MaurerEtal99} for the moderate values of
junction conductances.

In Section \ref{sec:crossover} we consider qualitatively the case of
an (almost) open dot, when the conductance of one of the contacts is
close to $G_0$. In this case, the $G(V_g)$ function does not exhibit
sharp peaks. Nevertheless, the question regarding the fluctuations of
the distance between the conductance maxima is still valid. We study
the origin of these fluctuations and estimate their variance.
Furthermore, we extend the consideration to the region of intermediate
values of the junction conductances. In this region, the
peak-to-valley ratio of the $G(V_g)$ dependence is large, but peaks
are not sharp (their width exceeds the level spacing in the dot). Thus
the developed theory yields understanding of the evolution of the peak
spacing fluctuations with the strength of the dot-lead coupling,
ranging from weak tunneling to weak reflection in the contacts
between the dot and the leads.

In this paper, we base our statistical description of the electron
states in the dot on the random matrix theory.  Despite the apparent
failure of RMT in explaining the peak spacing fluctuations in a closed
dot, this is an adequate approach for our purposes, because we
concentrate on the contribution to the peak spacing fluctuations
coming from the non-zero conductance of the contacts. This
contribution is determined by the randomness of the wave functions
near the contacts.  Fluctuating conductance peak heights, which depend
on the same random quantity, apparently are well described by
RMT.\cite{JalabertEtal92,ChangEtal96} 

In a partially open dot, the structure of the electron wave functions
in the vicinity of the contacts affects both the amplitudes and
positions of the conductance peaks. This results in a specific
cross-correlation between these two characteristics of the random
function $G(V_g)$. The correlation grows with the conductance of the
contacts; we calculate the corresponding correlation function.


\section{Conductance peak positions}
\label{sec:posit}
\subsection{RMT Hamiltonian}
\label{sec:RMTH}
In our consideration, the quantum dot is described by the Hamiltonian
of the constant interaction model:
\begin{equation}
\hat{H}_0=\sum_k \varepsilon_k^{\phantom{\dagger}}
 c^\dagger_k c^{\phantom{\dagger}}_k+
\frac{E_C}{2}\left(\sum_k c^\dagger_k c^{\phantom{\dagger}}_k - {\cal
N}\right)^2\;.
\label{H}
\end{equation}
In the framework of this model, the Coulomb repulsion of the electrons
in the dot is accounted for by a relatively simple charging term
[second term in (\ref{H})], and is characterized by only one
parameter, namely, the ``charging energy'' of the dot, $E_C\equiv
e^2/C$, where $C$ is the total capacitance of the dot. The gate
voltage $V_g$ is represented by dimensionless parameter ${\cal
 N}\equiv C_g V_g/e$, with $C_g$ being the capacitance of the dot
with respect to the gate. The (random) one-electron spectrum in the
dot is represented by the first term in Eq.~(\ref{H}), the energies
$\varepsilon_k$ are measured from the bottom of the band.

The weak dot-lead coupling can be accounted for by the tunneling
Hamiltonian
\begin{equation}
\hat{H}_t=\sum_{k,p}\left(t^{\phantom{\dagger}}_{kp}c^\dagger_{k}
c^{\phantom{\dagger}}_{p}+{\rm H.c.}\right)
+\sum_{k,q}\left(t^{\phantom{\dagger}}_{kq}c^\dagger_{k}
c^{\phantom{\dagger}}_{q}+{\rm H.c.}\right)
\;.
\label{Ht}
\end{equation}
Here indices $p$ and $q$ denote the states in the left and right lead
respectively.  The dimensionless conductance of the
dot-lead contacts is given by
\begin{eqnarray}
  \label{Gdef}
g_{L}&=& (2\pi)^2 \langle
|t_{kp}|^2\rangle\frac{\nu_{L}}{\Delta}\;, \;\nonumber\\
g_{R}&=& (2\pi)^2  \langle
|t_{kq}|^2\rangle\frac{\nu_{R}}{\Delta}\;, 
\end{eqnarray}
where $\langle...\rangle$ denotes the averaging over a statistical
ensemble, $\Delta\equiv\langle\varepsilon_{n+1}-\varepsilon_n\rangle$
is the average level spacing in the dot, and $\nu_\alpha$
($\alpha=L,R$) is the average density of states in the leads. A
reflectionless dot-lead channel corresponds to $g=1$; the conductance
of such a channel equals $G_0$ ($G_0/2$ for spinless fermions).

The tunneling matrix elements are proportional to the values of the
electron wave functions at the points ${\bf r}_\alpha$ of the dot-lead
contacts. The randomness of the wave functions results in the
fluctuations of the tunneling matrix elements in the
Hamiltonian~(\ref{Ht}). The fluctuations of the wave functions in
macroscopic leads can be neglected, and the tunneling matrix elements
can be written in the form
\begin{eqnarray}
t_{kp}&=& \frac{1}{2\pi}\sqrt{\frac{\Delta}{\nu_{L}}
\vphantom{\frac{G_{L,R}}{G_0}}}\,
\sqrt{g_L} \,
\frac{\psi^*_k({\bf r}_{L})}{\sqrt{\langle\psi_k^2\rangle}}\;,
\nonumber\\
t_{kq}&=& \frac{1}{2\pi}\sqrt{\frac{\Delta}{\nu_{R}}
\vphantom{\frac{G_{L,R}}{G_0}}}\,
\sqrt{g_R} \,
\frac{\psi^*_k({\bf r}_{R})}{\sqrt{\langle\psi_k^2\rangle}}\;.
\label{tkpi}
\end{eqnarray}

In Sec.~\ref{sec:positions}--\ref{sec:RG} we consider the model case of
spinless fermions, which is easier to follow. The proper extension
onto the case of real electrons with spins is performed in
Sec.~\ref{sec:spin}.
Since at high temperatures the mesoscopic fluctuations are washed out,
 in this paper we concentrate on the low-temperature limit, $T<\Delta$.


\subsection{Conductance peaks for spinless fermions}
\label{sec:positions}

Near a Coulomb blockade peak, where conductance through the dot is
mediated by one resonant level, the transmission coefficient of the
dot at the Fermi level is given by the Breit-Wigner formula
\begin{equation}
{\sf T}({\cal N})=\frac{\Gamma_L\Gamma_R}{\epsilon({\cal N})^2+
  [(\Gamma_L+\Gamma_R)/2]^2}\;, 
\label{BW}
\end{equation}
where $\Gamma_\alpha$ is the width of the resonant level with respect
to electron tunneling to the $\alpha$-th lead, and $\epsilon({\cal
  N})$ is the deviation from the resonance. The center of the peak in
the zero-bias conductance $G({\cal N})$ for spinless fermions
corresponds to such value of ${\cal N}$ that $\epsilon=0$. 

For an almost isolated dot ($t_{kp},t_{kq}\to 0$), function
$\epsilon({\cal N})$ near the $n$-th peak is given by
\begin{equation}
\epsilon=E^{(0)}_{n}({\cal N})-E^{(0)}_{n-1}({\cal N})\;.
\label{epsilon0}
\end{equation}
Here $E^{(0)}_n({\cal N})$ is the ground state energy of an isolated
dot with $n$ particles [Hamiltonian (\ref{H})]. The values of ${\cal
 N}$, at which the conductance has a peak are thus determined by the
equation
\begin{equation}
E^{(0)}_{n-1}\left({\cal N}^{(0)}_n\right)
=E_n^{(0)}\left({\cal N}^{(0)}_n\right)\;.
\label{N}
\end{equation}
Therefore, for an almost isolated dot, when $\epsilon$ is given by
Eq.~(\ref{epsilon0}), the spacing between the peaks is
\begin{equation}
{\cal U}_n^{(0)}\equiv{\cal N}^{(0)}_{n+1}-{\cal N}_{n}^{(0)}=
1+\frac{\varepsilon_{n+1}-\varepsilon_{n}}{E_C}\;.
\label{U}
\end{equation}
In this limit, the random component of the peak spacing has
the same statistics as the level spacings
$\Delta_n\equiv\varepsilon_{n+1}-\varepsilon_{n}$.

Upon the conductance increase, the electron transport through the dot
is still dominated by the tunneling via the resonant level, as long as
the effective level width remains small compared to the level spacing.
The parameters entering in Eq.~(\ref{BW}), however, acquire
corrections due to the virtual processes involving other levels in the
dot. The peak position is affected only by the corrections to
$\epsilon$. At weak dot-lead coupling, the latter can be evaluated in
the second order of the perturbation theory in $t_{kp}$, $t_{kq}$, so
$\epsilon$ is given by
\begin{eqnarray}
\epsilon&=&\left[E^{(0)}_{n}({\cal N})+\delta E'_{n}({\cal N})\right]
\nonumber\\
&&\qquad-\left[E^{(0)}_{n-1}({\cal N})+\delta E'_{n-1}({\cal N})\right]\;,
\label{epsilonprim}\\
\delta E'_n({\cal N})&=&
-\sum_{p}\left\{\sum_{-\infty}^{k=n-1}\!
\frac{|t_{kp}|^2\theta(\xi_{p})}{\xi_{p}\!+\!\varepsilon_{n}\!
-\!\varepsilon_k\!+\!E_C(n\!-\!{\cal N}\!-\!1/2)}\right.\nonumber\\
&&\left.
\quad+\sum^{\infty}_{k=n+1}\frac{|t_{kp}|^2\theta(-\xi_{p})}
{-\xi_{p}\!-\!\varepsilon_{n}\!+\!\varepsilon_k\! +\!E_C({\cal
    N}\!-\!n\!+\!3/2)}\right\} 
\nonumber\\
&&-\sum_q\left\{\ 
\vphantom{\sum_{-\infty}^{k=n-1}}
p\to q\ 
\right\}
\;,
\label{dEk}
\end{eqnarray}
where $\{p\to q\}$ denotes a term identical to the first term in
braces in Eq.~(\ref{dEk}), but with all $p$'s replaced with $q$'s.
Here $\delta E'_n$ is the correction to the energy of the state with
$n$ electrons in the dot due to the admixture of all the states with
$n\pm 1$ electron, except for the lowest-energy state with $n-1$
electrons. For spinless fermions, the mixing of the ground states with
$n$ and $n-1$ particles in the dot determines the width of the $n$-th
resonance (\ref{BW}) but does not affect its position.  The equation for
$\delta E'_{n-1}$ is similar to Eq.~(\ref{dEk}).

These corrections to $\epsilon$ shift the peak positions,
\begin{equation}
{\cal N}_n={\cal N}_n^{(0)}+\delta {\cal N}'_n\;.
\label{NNNNN}
\end{equation}
When the corrections $\delta E'_n({\cal N})$ are small, the peak
shifts $\delta{\cal N}'_n$ are also small and one can expand the right
hand side of Eq.~(\ref{epsilonprim}) in powers of $\delta{\cal N}$ in
the vicinity of ${\cal N}_n^{(0)}$. Then it follows from
Eqs.~(\ref{H}) and (\ref{epsilonprim}) that $\epsilon({\cal N})=0$ when
\begin{equation}
\delta {\cal N}'_n=
\left.\frac{
                \delta E'_n-\delta E'_{n-1}
           }
           {\frac{
                        \partial E^{(0)}_{n{\vphantom{-1}}}
                 }{
                        \partial {\cal N}
                 }
            -\frac{
                        \partial E^{(0)}_{n-1}
                  }{
                  \partial{\cal N}
                  }
            }
\right|_{{\cal N}={\cal N}^{(0)}_n}
=\left.
\frac{\delta E'_n-\delta E'_{n-1}}{E_C}\right|_{{\cal N}=
{\cal N}^{(0)}_n}\:.
\label{deltaN}
\end{equation}
Substituting the expressions for $\delta E'_{n-1}$ and $\delta E'_{n}$  
[Eq.~(\ref{dEk})] into Eq.~(\ref{deltaN}), and using Eq.~(\ref{tkpi}),
we find the shift of the conductance peak resulting from the dot-lead
coupling: 
\begin{eqnarray}
\delta{\cal N}'_n&=&\frac{1}{(2\pi)^2}\frac{\Delta}{E_C}
\left\{\sum_{\alpha=L,R}\sum_k
{\rm sign}\,(\varepsilon_k-\varepsilon_n)
g_\alpha
\right.
\nonumber\\
&&\qquad\qquad\left.
\vphantom{\sum_{-\infty}^{k=n-1}}
\vphantom{\sum_{\alpha=L,R}}
\times
\frac{|\psi_k({\bf r}_\alpha)|^2}{\langle|\psi_k|^2\rangle}
\ln\frac{E_C+|\varepsilon_n-\varepsilon_k|}{|\varepsilon_n-\varepsilon_k|}
\right\}\;.
\label{kvrhbi}
\end{eqnarray}
The ensemble average of this shift equals zero. However, the
mesoscopic fluctuations of the wave functions in the dot result in
non-zero random shifts of the conductance peaks in each particular
realization. Equation (\ref{kvrhbi}) yields the following expression
for the fluctuating component of the spacing between the $n$th and
$(n+1)$st conductance peaks:
\begin{eqnarray}
{\cal U}_n&=&{\cal U}^{(0)}_n
+\delta{\cal U}'_n\;,\label{main0}\\
\delta{\cal U}'_n&\equiv&\delta{\cal N}'_{n+1}-\delta{\cal N}'_{n}
\nonumber\\
&=&\frac{\Delta}{E_C}
\sum_{\parbox{0.5in}{\scriptsize
\quad$\alpha=L,R$\\
$l$=$n$,$n$+1}}\gamma'_\alpha\left(1-\frac{|\psi_l({\bf
r}_\alpha)|^2}{\langle|\psi_l|^2\rangle}\right)\;,
\label{main}
\end{eqnarray}
where
\begin{equation}
\gamma'_\alpha=\frac{1}{(2\pi)^2}g_\alpha
\ln\frac{E_C}{\Delta}\;.
\label{alpha}
\end{equation}

The term $\delta{\cal U}'_n$ in Eq.~(\ref{main0}) describes the
contribution of the dot-lead coupling to the fluctuations of the
spacing between the Coulomb blockade peaks. This contribution is
statistically independent of the term ${\cal U}^{(0)}_n$, which
accounts for the fluctuations in the energy spectrum of the dot and is
given by Eq.~(\ref{U}). The logarithmic factor in Eq.~(\ref{alpha})
is the result of summation over the lead states between the effective
upper and lower cut-offs, $E_C$ and $\Delta$. The contributions of the
wave functions of the states other than $n$ and $n+1$ do not contain
the large factor $\ln (E_C/\Delta)$ and are neglected.


\subsection{Spinless fermions: renormalization group approach}
\label{sec:RG}

Let us discuss the region of applicability of
Eqs.~(\ref{main})--(\ref{alpha}).  The right-hand side of
Eq.~(\ref{alpha}), obtained in the lowest non-vanishing order of the
perturbation theory in $t_{kp}$, $t_{kq}$, is proportional to the
product of the small parameter $g_\alpha$ and the large factor
$\ln (E_C/\Delta)$.  Examination of the higher orders of the
perturbation theory shows that the leading term of the $m$th order is
proportional to
\begin{equation}
\frac{\Delta}{E_C}\sqrt{g_\alpha}
\left[\sqrt{g_\alpha}\ln\frac{E_C}{\Delta}\right]^m \;.
\label{leading}
\end{equation}
Thus the results presented by Eqs.~(\ref{main})--(\ref{alpha}) are
valid only if the condition
\begin{equation}
\sqrt{g_\alpha}\ln\frac{E_C}{\Delta}\ll 1
\label{ptCondition}
\end{equation}
is satisfied. However, beyond the limits of the applicability of the
finite order perturbation theory, which are given by
Eq.~(\ref{ptCondition}), one can exploit the leading logarithm
approximation. It consists in summation of the leading terms [see
Eq.~(\ref{leading})] of all orders of the perturbation theory.  This
can be done by means of the same renormalization group technique,
which was initially developed by Anderson for the Kondo
problem.\cite{Anderson70} Application of this method to the Coulomb
blockade problem\cite{GlazmanMatveev89} allows one to obtain results
in the domain of parameters
\begin{equation}
\sqrt{g_\alpha}\ln\frac{E_C}{\Delta}\lesssim 1\;,
\label{RGCondition}
\end{equation}
which is wider than the region of applicability (\ref{ptCondition}) of
the finite order perturbation theory.  The technique used in
Ref.~\onlinecite{GlazmanMatveev89}, as well as the original technique
by Anderson, does not account for the mesoscopic fluctuations of the
one-electron wave functions. For our purposes, we need to extend the
renormalization group method to account for the randomness of the
electron states in the dot. Below we present the proper generalization
valid for a dot obeying the RMT statistics.

The Anderson's scaling procedure consists in transformation of the
initial Hamiltonian $\hat{H}_0+\hat{H}_t$ to a new one,
with smaller bandwidth and with the matrix elements renormalized to
compensate for this band reduction. The transformation operates within
the invariant space of Hamiltonians having generic form
\begin{equation}
\sum_j
\left[{\sf E}^{(j)}_0+\sum_{\eta_1\eta_2}
{\sf T}_{\eta_1\eta_2}^{(j)}
c^\dagger_{\eta_1}c^{\phantom{\dagger}}_{\eta_2}
\right]\hat{P}_j\;,
\label{HT}
\end{equation}
where the indices $\eta_1,\eta_2$ run through the one-particle states
in the dot and both leads, and $\hat{P}_j$ is the operator of projection to
the Hilbert subspace of states with $j$ particles in the dot. 
Near the $n$-th peak, the renormalization of the Hamiltonian occurs
mostly due to the virtual transitions between the states with $n$ and
$n-1$ particles in the dot, and it is sufficient to limit our
consideration to these two sets of states. Then the summation in
Eq.~(\ref{HT}) must be done over $j=n-1,n$ only, and ${\sf
  T}_{kp}^{(n)},\ {\sf T}_{pk}^{(n-1)},\ {\sf
  T}_{kq}^{(n)},\ {\sf T}_{qk}^{(n-1)}\equiv 0$ because the
corresponding transitions involve states outside this subspace of
states.

The transformation starts from Hamiltonian (\ref{H})--(\ref{Ht})
operating within the electron band of width $D_0\sim E_C$. In the
course of renormalization, the band width $D$ is reduced in steps
starting from $D=D_0$. As we mentioned above, the matrix elements must
be adjusted to compensate for the band reduction: after each step, the
transition amplitudes calculated with the renormalized Hamiltonian
$\hat{\sf H}(D-d D)$ must coincide with those derived from the
Hamiltonian $\hat{\sf H}(D)$ of the previous step. In the leading
logarithmic approximation, which we employ, these amplitudes should be
calculated in the second order of the perturbation theory. Then the
scaling law, which dictates how a matrix element must be changed when
the bandwidth is reduced from $D$ to $D-d D$, in general reads
\begin{equation}
d{\sf T}^{(j)}_{\eta_1\eta_2}=\sum_{D-d D<|\xi_{\eta_3}|<D}
\frac{{\sf T}^{(j_3)}_{\eta_1\eta_3}{\sf T}^{(j)}_{\eta_3\eta_2} 
}{-{\cal E}[\eta_2\to\eta_3]}\;,
\label{Tscaling}
\end{equation}
where ${\cal E}[\eta_2\to\eta_3]$ is the energy difference associated
with the (virtual) transition $\eta_2 \to \eta_3$,
and $j_3$ is the number of electrons in the dot after this transition.
Here we present explicitly only the equation for ${\sf T}^{(n)}_{k_1 k_2}$:
\begin{eqnarray}
&&d{\sf T}_{k_1 k_2}^{(n)}=\!\!
\sum_{-D<\xi_{p}<-D+d D}\!\!
\frac{{\sf T}_{k_1 p}{\sf T}_{pk_2}}{D}
+\sum_{-D<\xi_{q}<-D+d D}\!\!
\frac{{\sf T}_{k_1 q}{\sf T}_{qk_2}}{D}
\nonumber\\
&&
-\!\!\sum_{D-d D<\xi_{k_3}<D}\!\!
\frac{{\sf T}^{(n)}_{k_1k_3}{\sf T}^{(n)}_{k_3k_2}}{D}
+\!\!\sum_{-D<\xi_{k_3}<-D+d D}\!\!
\frac{{\sf T}^{(n)}_{k_1k_3}{\sf T}^{(n)}_{k_3k_2}}{D}\;.
\label{Tkk0}
\end{eqnarray}
The equations for the other parameters of the Hamiltonian have similar
form.  The renormalization procedure also generates the constant term
${\sf E}_0$:
\begin{equation}
d{\sf E}^{(j)}_0=-\!\!\!\sum_{\parbox{1.05in}{\scriptsize
$-D\!<\!\xi_{\eta_1}\!\!<\!-D+d D$
\nonumber\\
$D-d D<\xi_{\eta_2}<D$
}}\!\!\!\!\!
\frac{{\sf T}^{(j_1)}_{\eta_2\eta_1}{\sf T}^{(j)}_{\eta_1\eta_2}}{2D}
+\!\!\!\sum_{-D<\xi_{\eta}<-D+d D} \!\!{\sf T}_{\eta\eta}^{(j)}\;.
\label{Escaling}
\end{equation}
The initial Hamiltonian $\hat{\sf H}(D=E_C)\equiv\hat{H}_0+\hat{H}_t$
[Eqs.~(\ref{H})--~(\ref{Ht})] in terms of Eq.~(\ref{HT}) with
$j=n-1,n$ corresponds to
\begin{eqnarray}
{\sf E}_0^{(j)}&=&0\;,\nonumber\\
{\sf T}_{kp}&\equiv&{\sf T}_{pk}^*=t_{kp}\:,\;
{\sf T}_{kq}\equiv{\sf T}_{qk}^*=t_{kq}\:,\;
\label{RGinit}\\
{\sf T}^{(j)}_{k_1k_2}&=&0\:,\;
{\sf T}^{(j)}_{p_1p_2}=0\:,\;{\sf T}^{(j)}_{q_1q_2}=0\:.\;\nonumber
\end{eqnarray}
It is random because of the fluctuations of one-electron wave
functions [see Eq.~(\ref{tkpi})] and level spacings. The 
Hamiltonian $\hat{\sf H}(D<E_C)$, which is produced by the scaling
transformation from $\hat{H}_0+\hat{H}_t$, is therefore also random.

As it can be seen from Eq.~(\ref{Tkk0}), the randomness of the initial
Hamiltonian $\hat{H}_0+\hat{H}_t$ is inherited by the renormalized
Hamiltonian $\hat{\sf H}(D)$ in two ways. First, there are random
contributions from the states $k_1$ and $k_2$.  Second, there are also
random contributions from the states $k_3$, being eliminated by the
band reduction.  The final renormalized matrix element ${\sf T}_{k_1k_2}$
will still be proportional to the (random) wave functions
$\psi^*_{k_1}$ and $\psi^{\phantom{*}}_{k_2}$. On the contrary, the random
contributions from the eliminated states $k_3$ are added together and
their (independent) fluctuations are averaged out. This consideration
leads one to the conclusion that the solutions of the RG equations
(\ref{Tscaling}) can be cast in the form:
\begin{mathletters}
\label{lmn1}
\begin{eqnarray}
{\sf T}_{k_1k_2}^{(j)}(D)&=&\pm
\Delta
\sum_{\alpha=L,R}
\frac{\psi_{k_1}^*({\bf r}_\alpha)\psi_{k_2}^{\phantom{*}}({\bf
    r}_\alpha)}{\langle\psi_k^2\rangle} 
\,\lambda_\alpha(D)\;,\label{_tkk}\\
{\sf T}_{kp}(D)&=&\sqrt{\frac{\Delta}{\nu_{L}}}
\frac{\psi_k({\bf r}_{L})}{\sqrt{\langle\psi_k^2\rangle}}
\,\mu_{L}(D)\;,\label{_tkp}\\
{\sf T}_{kq}(D)&=&\sqrt{\frac{\Delta}{\nu_{R}}}
\frac{\psi_k({\bf r}_{R})}{\sqrt{\langle\psi_k^2\rangle}}
\,\mu_{R}(D)\;,\label{_tkq}\\
{\sf T}_{p_1p_2}^{(j)}(D)&=&\mp\frac{1}{\nu_{L}}
\,\lambda_{L}(D)\;.\label{_tpp}\\
{\sf T}_{q_1q_2}^{(j)}(D)&=&\mp\frac{1}{\nu_{R}}
\,\lambda_{R}(D)\;.\label{_tqq}
\end{eqnarray}
\end{mathletters}
Here the upper sign in the right-hand-sides corresponds to $j=n-1$ and
the lower sign corresponds to $j=n$.  From
Eqs.~(\ref{Tscaling}),~(\ref{RGinit}), and (\ref{lmn1}), we derive the
scaling equations for the non-fluctuating parameters $\lambda_i$ and
$\mu_i$, which determine the behavior of the Hamiltonian in the course
of the scaling procedure:
\begin{equation}
\frac{d\lambda_\alpha(D)}{d\ln D}=\mu_\alpha^2(D)\;,\qquad
\frac{d\mu_\alpha(D)}{d\ln D}=4\lambda_\alpha(D)\mu_\alpha(D)\label{lmnscaling}
\end{equation}
with the initial conditions
\begin{equation}
\lambda_\alpha(D=E_C)=0\;,\qquad\mu_\alpha(D=E_C)=
\frac{1}{2\pi}\sqrt{g_\alpha}\;.
\end{equation}
They coincide with those derived in
Ref.~\onlinecite{GlazmanMatveev89} and have solutions
\begin{eqnarray}
\lambda_\alpha(D)&=&
\frac{1}{4\pi}\sqrt{g_\alpha}
\tan\left[\frac{1}{\pi}\sqrt{g_\alpha}\ln\frac{E_C}{D}
\right]\;,\nonumber\\
\mu_\alpha(D)&=&\frac{1}{2\pi}\sqrt{g_\alpha}
\left\{\cos\left[\frac{1}{\pi}\sqrt{g_\alpha}\ln\frac{E_C}{D}
\right]\right\}^{-1}\;.
\label{lambdamufinal}
\end{eqnarray}

The energy shift ${\sf E}_0$ also accumulates in the course of
renormalization. Its random part, which stems from the fluctuations of
the wave functions of the eliminated states $k_3$, averages out due to
the summation over a large number of these intermediate states,
similarly to the corresponding contribution to the matrix elements
${\sf T}_{p_1p_2}$ and ${\sf T}_{q_1q_2}$.

The transformation can proceed until the band width is reduced to the
value of the order of the mean level spacing $\Delta$.  After the
renormalization, the ground state energy can be found with the help of
Hamiltonian $\hat{\sf H}(D\sim\Delta)$ as the sum of ${\sf E}_0$
and the correction due to the dot--lead tunneling in the reduced band.
As we explained above, the latter term is the main source of
fluctuations of the ground state energy. It can be calculated in the
second order of the perturbation theory in ${\sf T}_{\eta_1\eta_2}$,
similarly to Eq.~(\ref{dEk}).  This calculation finally yields
Eq.~(\ref{main}) with
\begin{equation}
\gamma'_\alpha=\frac{1}{4\pi}\sqrt{g_\alpha}
\tan\left[\frac{1}{\pi}\sqrt{g_\alpha}\ln\frac{E_C}{\Delta}
\right]\;.
\label{alphaRG}
\end{equation}
If the argument of the tangent function is much less than unity,
Eq.~(\ref{alphaRG}) is reduced to Eq.~(\ref{alpha}), as expected.


\subsection{Electrons with spin}
\label{sec:spin}

In this section we will refer to the same Hamiltonian
(\ref{H})--(\ref{Ht}), with summation over the spin index added.  The
conductance peaks are numbered as follows: the $(2m-1)$-th conductance
peak corresponds to the resonant tunneling of an electron through the
empty $m$-th level, and the $(2m)$-th peak corresponds to resonant
electron tunneling via the $m$-th level already occupied by one
electron.

In principle, the spin degeneracy can lead to formation of the
many-body state, which enhances the conductance in the valley between
the $(2m-1)$-th and $(2m)$-th peaks, so Eq.~(\ref{BW}) is no longer
valid. This phenomenon is commonly referred to as the Kondo effect.
However if the temperature is larger than the widths of the dot
levels, which is the case under consideration, the formation of this
many-body state is suppressed, and we can employ the rate equation
formalism to evaluate the conductance of the dot. This approach yields
the following expression for the dot conductance:\cite{Beenakker91}
\begin{equation}
  \label{spincond}
  G({\cal N})=G_0\frac{g_L g_L}{g_L+g_R}
\frac{\partial f_F[\epsilon({\cal N})]/\partial\epsilon}
     {1+f_F[\pm\epsilon({\cal N})]}\;,
\end{equation}
where $f_F(\xi)\equiv 1/[\exp(\xi/T)+1]$ is the Fermi distribution
function. The sign in the denominator should be taken ``$+$'' for even
peaks and ``$-$'' for odd peaks. One can see that the peaks in the
conductance $G({\cal N})$ are shifted from the points of charge
degeneracy, which correspond to $\epsilon({\cal N})=0$, by
$\pm (\ln 2/2)T$, which results in the correction to the peak spacings:
   \begin{equation}
     \label{UT}
     \delta{\cal U}_{2m-1}^{(T)}=(\ln2)\frac{T}{E_C}\;,\quad
     \delta{\cal U}_{2m}^{(T)}=-(\ln2)\frac{T}{E_C}\;.
   \end{equation}
  
  For an almost isolated dot, the distances between the charge
  degeneracy points are given by
\begin{mathletters}
  \label{uspinisol}
\begin{eqnarray}
  {\cal U}^{(0)}_{2m-1}&=&1\;,
\label{uspinisolodd}\\
  {\cal U}^{(0)}_{2m}&=&
  1+\frac{\varepsilon_{m+1}-\varepsilon_m}{E_C}\;.
\label{uspinisoleven}
\end{eqnarray}
\end{mathletters}

To obtain the expressions for the conductance peak spacings for a
partially open dot, one has to add up ${\cal U}^{(0)}$, $\delta{\cal
  U}^{(T)}$, and the contribution to the peak spacing due to the
finite dot--lead tunneling. This contribution has two components.  

The first component, $\delta{\cal U}'_n$, is analogous to the one
calculated in Sec.~\ref{sec:positions}, it comes from the transitions
involving all levels in the dot except for the resonant one.  The
spacing ${\cal U}_{2m}$ is between the conductance peaks corresponding
to filling of two consecutive orbital levels. The result for
spinless fermions is given by Eq.~(\ref{main}). For electrons with
spin the corresponding correction acquires an additional factor of 2:
\begin{mathletters}
  \label{Uspinprim}
  \begin{equation}
    \label{Uspinprimeven}
\delta{\cal U}'_{2m}=2\frac{\Delta}{E_C}
\sum_{\parbox{0.5in}{\scriptsize
\quad$\alpha=L,R$\\
$l$=$m$,$m$+1}}
\gamma'_\alpha
\left(1-\frac{|\psi_l({\bf
      r}_\alpha)|^2}{\langle|\psi_l|^2\rangle}\right)\:.
  \end{equation}
The correction of this type contribute equally to the positions ${\cal
  N}_{2m-1}$, ${\cal N}_{2m}$
or the peaks, corresponding to the same orbital level.
As the result
  \begin{equation}
    \label{Uspinprimodd}
\delta{\cal U}'_{2m-1}=0\:.
  \end{equation}
\end{mathletters}

The other component, $\delta{\cal U}''_n$, is due to the virtual
transitions from/to the resonant level.  Let us consider, for
definiteness, the $2m$-th conductance peak, whose position is
determined by the energies of the ground states $|2m-1\rangle$ and
$|2m\rangle$, with $2m-1$ and $2m$ electrons in the dot respectively.
For spinless fermions, virtual transitions involving the $m$-th level
do not lead to the shift of the peak position, because of the
electron-hole symmetry. The spin degeneracy, however, removes this
symmetry and as the result the shifts of the ground state energies
$\delta E_{2m}$ and $\delta E_{2m-1}$ are not equal:\cite{Haldane78}
The state $|2m\rangle$ acquires an admixture of the excited states
with $2m-1$ electrons in the dot, and the corresponding correction is
given by
\begin{mathletters}
\label{dEkspin}
  \begin{equation}
    \label{dEk0}
\delta E''_{2m}=-2 \left[\sum_{p}
\frac{|t_{mp}|^2 \theta(\xi_{p})}{\xi_{p}}
+\sum_{q}
\frac{|t_{mq}|^2 \theta(\xi_{q})}{\xi_{q}}\right]\;,
\end{equation}
where the prefactor 2 is due to the spin degeneracy.  The state
$|2m-1\rangle$ intermixes with the excited states having $2m$ or
$2m-2$ electrons in the dot. At the point of the $2m$-th
conductance peak, which we consider, the states $|2m-1\rangle$ and
$|2m\rangle$ have equal energy, but the energy of the state
$|2m-2\rangle$ is higher, because of the charging. Therefore the
coupling of the states with $2m-1$ and $2m-2$ electrons in the dot is
suppressed and the correction $\delta E''_{2m-1}$ is smaller than
$\delta E''_{2m}$:
\begin{eqnarray}
  \label{eq:dEk1}
\delta E''_{2m-1}&=&-\left[\sum_{p}
\frac{|t_{mp}|^2 \theta(-\xi_{p})}{-\xi_{p}}
+ \sum_{p}
\frac{|t_{mp}|^2 \theta(\xi_{p})}{\xi_{p}+2E_C}  \right]\nonumber\\
&&-\left[
\vphantom{\frac{|t_{mp}|^2 \theta(-\xi_{p})}{-\xi_{p}}}
p\to q
\right]\;.
\end{eqnarray}
\end{mathletters}
Substitution of Eqs.~(\ref{dEkspin}) into Eq.~(\ref{deltaN}) yields
\begin{equation}
\delta{\cal N}''_{2m}=-\frac{\Delta}{E_C}
\sum_{\alpha=L,R}\gamma''_\alpha
\frac{|\psi_m({\bf r}_\alpha)|^2}{\langle|\psi_m|^2\rangle}\;,
\label{deltaNHaldane}
\end{equation}
with
\begin{equation}
  \label{alphaprimprim}
\gamma''_ \alpha=\frac{1}{(2\pi)^2}g_\alpha
\ln\frac{E_C}{T}\;.
\end{equation}
The shift of an odd peak can be calculated analogously and is given by
\begin{equation}
  \label{ohoho}
  \delta{\cal N}''_{2m-1}=-\delta{\cal N}''_{2m}\;.
\end{equation}

In contrast to the correction from the non-resonant levels
[Eq.~(\ref{deltaN})], the correction $\delta{\cal N}''_n$ has a
non-zero ensemble average,
\begin{equation}
\langle\delta{\cal N}''_n\rangle 
=(-1)^{n-1}\sum_{\alpha=L,R}\gamma''_\alpha\frac{\Delta}{E_C}\;.
\label{Nspinav1}
\end{equation}

Equations (\ref{deltaNHaldane}),~(\ref{ohoho}) yield
\begin{mathletters}
\label{Uspinprimprim}
\begin{eqnarray}
&&\delta{\cal U}''_{2m}=\frac{\Delta}{E_C}\sum_{\parbox{0.5in}{\scriptsize
\quad$\alpha=L,R$\\
$l$=$m$,$m$+1}}
\gamma''_\alpha\frac{|\psi_l({\bf r}_\alpha)|^2}{\langle|\psi_l|^2\rangle}\;,
\\
&&\delta{\cal U}''_{2m-1}=-2\frac{\Delta}{E_C}\sum_{\alpha=L,R}
\gamma''_\alpha 
\frac{|\psi_{m}({\bf r}_\alpha)|^2}{\langle|\psi_{m}|^2\rangle}\;.
\end{eqnarray}
\end{mathletters}

The full expression for the peak spacings in the case of electrons
with spin is 
\begin{equation}
{\cal U}_n={\cal U}^{(0)}_n+\delta{\cal U}^{(T)}_n+
\delta{\cal U}'_n+\delta{\cal U}''_n\;,
\label{finalspin}
\end{equation}
where $\delta{\cal U}_n^{(T)}$ and ${\cal U}_n^{(0)}$ are given by
Eqs.~(\ref{UT}) and (\ref{uspinisol}) respectively.

Equations (\ref{deltaNHaldane})-(\ref{alphaprimprim}), derived with
the help of the second order perturbation theory in $t_{kp}$,
$t_{kq}$, are valid only at relatively small values of the
conductances $g_\alpha$ of the contacts (see Sec.~\ref{sec:RG}). When
the perturbation theory fails, the proper expression for
$\gamma''_\alpha$ can be obtained with the help of the renormalization
group method, described in Sec.~\ref{sec:RG}. If the temperature is of
the order of the mean level spacing, then
the RG transformation must be stopped when the bandwidth $D$ is reduced to
$\Delta$; the renormalized $\gamma''_\alpha$ equals
$\gamma'_\alpha$, which is given by Eq.~(\ref{alphaRG}).

If $T<\Delta$, then we
have to continue the renormalization procedure to the bandwidths
$D<\Delta$.  There is only one level in the dot within such a band, 
so Eqs.~(\ref{Tscaling}), (\ref{lmn1}) yield:
\begin{equation}
  \label{lprimprim}
  \frac{d\lambda_\alpha(D)}{d\ln D}=\mu_\alpha^2(D)\;,\qquad
  \frac{d\mu_\alpha(D)}{d\ln D}=2\mu_\alpha(D)\lambda_\alpha(\Delta)\;.
\end{equation}
The solutions of these equations determine the renormalization of the
tunneling matrix elements ${\sf T}_{k_1k_2}$, ${\sf T}_{kp}$, and
${\sf T}_{kq}$ given by Eqs.~(\ref{_tkk})--(\ref{_tkq}). As for ${\sf
  T}_{p_1p_2}$ and ${\sf T}_{q_1q_2}$, they are not renormalized anymore and
are given by Eqs.~(\ref{_tpp})--(\ref{_tqq}) with $D=\Delta$.  The
renormalization procedure should be carried out until $D\sim T$.
Finally we extend our result (\ref{deltaNHaldane}) to a wider region
of parameters, defined by Eq.~(\ref{RGCondition}), with parameter
$\gamma''_\alpha$ given by
\begin{equation}
  \label{RGalphaprimprim}
  \gamma''_\alpha=\gamma'_\alpha+
\frac{\mu_\alpha^2(\Delta)}{2\lambda_\alpha(\Delta)}
\left[\left(\frac{\Delta}{T}\right)^{2\lambda_\alpha(\Delta)}-1\right]\;.
\end{equation}
Here $\lambda_\alpha(\Delta)$ and $\mu_\alpha(\Delta)$ are determined
by Eqs.~(\ref{lambdamufinal}). The first term in
Eq.~(\ref{RGalphaprimprim}) is produced in the course of 
renormalization while the band is reduced to $\Delta$; the second term
is produced by the band reduction from $\Delta$ to $T$.  Like in the
spinless case, Eq.~(\ref{RGalphaprimprim}) reduces to
Eq.~(\ref{alphaprimprim}) at small values of the dot-lead conductance,
{\em i.e.} when the condition (\ref{ptCondition}) is satisfied. 


\section{Statistical properties of the peak spacing}
\label{sec:statprop}
\subsection{Peak spacing fluctuations}

Now we discuss how the finite tunneling strength affects the
distribution function of the peak spacings, which is given by
\begin{eqnarray}
{\sf P}_n\left({\cal U}\right)&=&\int d{\cal U}_n^{(0)}
{\sf P}^{(0)}\left({\cal U}^{(0)}_n\right)
\prod_{\alpha=L,R}d\{\psi\}({\bf r}_\alpha)
{\sf P}\left(\{\psi\}({\bf r}_\alpha)\right)
\nonumber\\
&\times&\delta\left({\cal U}-{\cal U}_n\right)\;.
\end{eqnarray}
Here ${\cal U}_n$ is determined by Eq.~(\ref{finalspin}).  It follows
from Eq.~(\ref{uspinisol}) that the distribution function of ${\cal
  U}_n^{(0)}$, ${\sf P}^{(0)}$, is a $\delta$-function for $n=2m-1$, and
is given by the Wigner-Dyson distribution function for $n=2m$. The
set $\{\psi\}$ denotes the wave functions entering
Eqs.~(\ref{Uspinprim}) and (\ref{Uspinprimprim}) for $\delta{\cal
  U}'_n$ and $\delta{\cal U}''_n$.  The values $\psi_m({\bf r})$ are
distributed according to the Porter-Thomas statistics.

At small junction conductance, corrections $\delta{\cal U}'_n$,
$\delta{\cal U}''_n$ are negligible. In this limit, the distribution
of the peak spacings is manifestly bimodal. Odd values of $n$
correspond to the spacing between two peaks originating from filling
of the same orbital level in the dot; the corresponding distribution
function is infinitely sharp, and is centered at ${\cal U}=1+T\ln
2/E_C$.  Peak spacings at even $n$ correspond to filling of the
consecutive orbital levels, and therefore the distribution function for
them follows the Wigner-Dyson statistics. Note, that at a finite
temperature the centers of the ``even'' and ``odd'' distributions are
shifted towards each other, so these two distributions overlap. This
shift, which is the result of interaction between the electrons
tunneling via a spin-degenerate level, was not taken into account in
the analysis of experimental data.\cite{PatelEtal98}

At finite contact conductance, mesoscopic fluctuations of the wave
functions in the dot yield a finite width of the ``odd'' distribution
and broaden the ``even'' distribution. These effects stem from a
finite value of $\delta{\cal U}'_n$ and $\delta{\cal U}''_n$. The
resulting widths are characterized by
\begin{mathletters}
\label{varspin}
  \begin{eqnarray}
    \label{varodd}
\langle(\delta{\cal U}_{2m-1})^2\rangle&=&
\frac{8}{\beta}\sum_{\alpha=L,R}(\gamma''_\alpha)^2
\left(\frac{\Delta}{E_C}\right)^2\;,\\
    \label{vareven}
\langle(\delta{\cal U}_{2m})^2\rangle&=&
\frac{4}{\beta}\sum_{\alpha=L,R}(2\gamma'_\alpha-\gamma''_\alpha)^2
\left(\frac{\Delta}{E_C}\right)^2\nonumber\\
&+&\frac{\langle\delta\Delta_n^2\rangle}{E_C^2}\;.
  \end{eqnarray}
\end{mathletters}
Here $\beta=1$ and $2$ for orthogonal and unitary ensembles
respectively.  For smaller values of junction conductances $g_\alpha$
[such that condition (\ref{ptCondition}) is satisfied],
$\gamma'_\alpha$ and $\gamma''_\alpha$ are proportional to $g_\alpha$
and are given by Eqs.~(\ref{alpha}) and (\ref{alphaprimprim})
respectively.  Larger conductances cause renormalization of
$\gamma'_\alpha$ and $\gamma''_\alpha$; the resulting expressions for
them are given by Eqs.~(\ref{alphaRG}) and (\ref{RGalphaprimprim}).
 
One can see from Eq.~(\ref{vareven}) that there are two independent
additive contributions to the fluctuations of the peak spacing. The
first one is due to the fluctuations of the wave functions and is
proportional to the junction conductance. The second one comes from
the fluctuations of the electron spectrum. If the latter is described
by RMT, then $\langle\delta\Delta_n^2\rangle=[(4/\pi)-1]\Delta^2$ and
$\langle\delta\Delta_n^2\rangle=[(3\pi/8)-1]\Delta^2$ for orthogonal
and unitary ensembles respectively. The experimentally measured
\cite{PatelEtal98} $\langle\delta\Delta_n^2\rangle$ apparently are
larger than the values predicted by RMT. Regardless this discrepancy,
the contribution coming from the randomness of the wave functions can
be singled out from experimental data by comparing the values of
$\langle \delta{\cal U}_n^2\rangle$ measured at various junction
conductances.

The average spacing between the peaks is also affected by finite
conductance:
\begin{mathletters}
  \label{spinav}
\begin{eqnarray}
  \label{spinavodd}
  \langle{\cal U}_{2m-1}\rangle&=&1+(\ln 2)\frac{T}{E_C}
  -2\sum_{\alpha=L,R}\gamma''_\alpha\frac{\Delta}{E_C} \;,\\ 
  \langle{\cal U}_{2m}\rangle&=&1+\frac{\Delta}{E_C}-(\ln 2)\frac{T}{E_C}
  +2\sum_{\alpha=L,R}\gamma''_\alpha\frac{\Delta}{E_C}\;.
\end{eqnarray}
\end{mathletters}

\begin{figure}
  \epsfxsize=8.5cm \centerline{\epsfbox{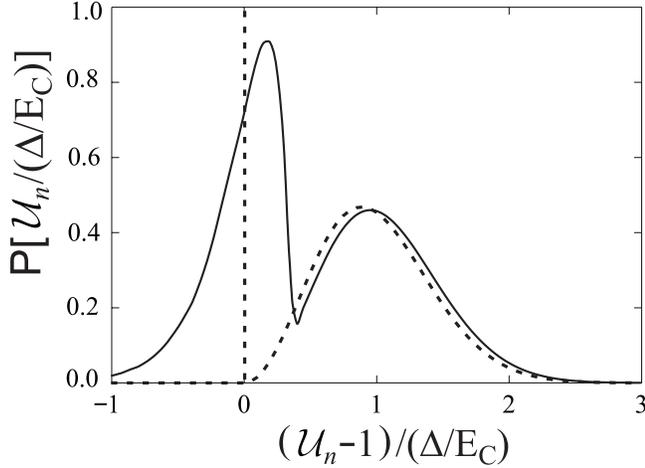}}
\caption{The cumulative distribution function of the ``even'' and
  ``odd'' spacings between the Coulomb blockade peaks, 
  ${\cal U}_n/(\Delta/E_C)$, given by Eqs.~(\protect\ref{Uspinprim}), 
(\protect\ref{Uspinprimprim}) for
  $\gamma'_\alpha=0.075$, $\gamma''_\alpha=0.1$, $T=0.5\Delta$ (solid line). 
The dotted line shows the ``bimodal'' distribution ${\sf P}^{(0)}$
  consisting of the Wigner-Dyson distribution (``even'' spacings) and
  a $\delta$-peak (``odd'' spacings). The shown distributions
  correspond to the Gaussian unitary ensemble.
\label{fig:Ucomp}
}
\end{figure}

One can also consider the distribution function averaged over even and
odd peak spacings. A sample plot of this function is shown in
Fig.~\ref{fig:Ucomp}. It is instructive, following the experimental
paper,\cite{PatelEtal98} to compare this function to the distribution
function ${\sf P}^{(0)}({\cal U})$ of the spacings between the charge
degeneracy points in an almost isolated dot. The ``bimodal''
distribution ${\sf P}^{(0)}({\cal U})$ consists of the Wigner-Dyson
distribution (``even'' spacings) and a $\delta$-peak (``odd''
spacings).  The key differences of ${\sf P}({\cal U})$ from ${\sf
  P}^{(0)}({\cal U})$ are as follows: (i) the $\delta$-peak
corresponding to odd spacings is broadened (ii) the centers of the
distribution maxima corresponding to even and odd peak spacings are
shifted by $\pm[(\ln 2)T/E_C-2(\sum_\alpha\gamma''_\alpha)\Delta/E_C$]
(iii) unlike ${\sf P}^{(0)}({\cal U})$, the distribution ${\sf
  P}({\cal U})$ has a tail which extends into the region ${\cal U}<1$.

\narrowtext

One can see from Eq.~(\ref{varspin}) that the fluctuations of wave
functions and of the level spacings contribute equally to the width of
${\sf P}({\cal U})$ at $\gamma'_\alpha\sim\gamma''_\alpha\sim 0.15$.
The dimensionless parameters $\gamma'_\alpha$ and $\gamma''_\alpha$
depend not only on the junction conductances, but also on the ratios
$E_C/\Delta$ and $E_C/T$, see Eqs.~(\ref{alphaRG}) and
(\ref{RGalphaprimprim}).  Using the values $E_C=300 \mu eV$, $\Delta=7
\mu eV$, and $T\sim\Delta$ from the experiment of
Ref.~\onlinecite{PatelEtal98}, we obtain that the values
$\gamma'_\alpha\sim 0.05$ are reached when the junction conductances
equal to $0.5 G_0$.  Thus a considerable contribution of wave function
fluctuations to the randomness of the interpeak spacings can be
achieved when the conductance peaks are still well-defined.


\subsection{Correlations between the heights of the peaks and their
  spacings} 

We have shown that for a partially open dot, the conductance peak
positions depend on the values of the one-electron wave functions at
the points of the dot-lead contacts. On the other hand, these values
determine also the conductance peak amplitudes:
\begin{equation}
  G_{2m-1}^{\rm (max)}=
  G_{2m}^{\rm (max)}
\propto 
  \frac{g_L|\psi_m({\bf r}_L)|^2 g_R|\psi_m({\bf r}_R)|^2}
  {g_L|\psi_m({\bf r}_L)|^2+ g_R|\psi_m({\bf r}_R)|^2}\;.
\label{G}
\end{equation}
Therefore the spacing between two peaks must be correlated with the
heights of these peaks: the higher the peaks are, the less (on
average) is the spacing between them.

The joint distribution function of the conductance peak height
$G^{\rm (max)}$ and the spacing of this peak from its
neighbors $\tilde{\cal U}$ is given by
\begin{eqnarray}
{\sf P}\left(\tilde{\cal U},G^{\rm (max)}\right)&=&
\int d{\cal U}_n^{(0)}
{\sf P}^{(0)}\left({\cal U}^{(0)}_n\right)\nonumber\\
&&\times
\prod_{\alpha=L,R}d\{\psi\}({\bf r}_\alpha)
{\sf P}\left(\{\psi\}({\bf r}_\alpha)\right)
\nonumber\\
&&\times\delta\left(\tilde{\cal U}-\tilde{\cal U}_n\right)
\delta\left(G^{\rm (max)}-G^{\rm (max)}_n\right)\;.
\label{joint}
\end{eqnarray}
Here the spacing of the $n$-th peak from its neighbors is
\begin{displaymath}
\tilde{\cal U}_n\equiv\frac{{\cal U}_n+{\cal U}_{n+1}}{2}\;,  
\end{displaymath}
and ${\cal U}_n$ is given by Eq.~(\ref{finalspin}).  The
distribution ${\sf P}(\tilde{\cal U},G^{({\rm max})})$ is skewed, {\em
  i.e.}  $\langle\delta \tilde{\cal U}\, \delta G^{({\rm
    max})}\rangle\neq 0$, because of the dependence of $\tilde{\cal U}$ on
$\{\psi\}$.  The dimensionless cross-correlation parameter grows with
junction conductance,
\begin{eqnarray}
K_{G-{\cal U}}&\equiv&
\frac{\langle \delta G_n^{\rm (max)} \delta \tilde{\cal U}_n\rangle}
{\langle G_n^{\rm (max)}\rangle (\Delta/E_C)}\nonumber\\
&=&-\sum_{\parbox{0.5in}{\scriptsize
$\alpha=L,R$\\
$\alpha'\neq \alpha$}}
{\cal F}\left(\frac{g_{\alpha'}}{g_{\alpha}}\right)
\left(2\gamma'_\alpha+\gamma''_\alpha\right)\;.
\label{corr}
\end{eqnarray}
Function ${\cal F}(x)>0$ is given by
\begin{displaymath}
  {\cal F}(x)=\frac{2\sqrt{x}}{1+\sqrt{x}}
\end{displaymath}
for the orthogonal ensemble and by
\begin{displaymath}
  {\cal F}(x)=\frac{2x^3+3x^2-6x+1-6x^2\ln x}{(x-1)(x^2-1-2x\ln x)}-1
\end{displaymath}
for the unitary ensemble. In the case $g_L=g_R$, the argument of the
function ${\cal F}$ in Eq.~(\ref{corr}) equals unity; ${\cal F}(1)=1$
and $3/2$ for the orthogonal and unitary ensemble respectively.


\subsection{Fluctuations of spacing between distant peaks}
\label{sec:distant}
In the previous sections we considered only the fluctuations of
spacing between the neighboring peaks. Now we will discuss the
statistical properties of the spacing between the $n$-th and $(n+l)$-th
peaks,
\begin{equation}
  \label{distu}
{\cal U}_n(l)\equiv{\cal N}_{n+l}-{\cal N}_n\;, 
\end{equation}
with $l>1$. Here we must mention that in real dots adding electrons
apparently scrambles the one-electron states.\cite{PatelEtal98-2}
However, according to the experiment,\cite{MaurerEtal99,PatelEtal98-2}
the changes in the dot's one-electron spectrum and wave functions are
small when the number of added electrons is small. Under these
conditions, the same realization of the random parameters of the
Hamiltonian (1)-(2) determines the positions of $n$th and $(n+l)$th
peaks.  The average spacing between the peaks can be derived from
Eqs.~(\ref{UT}), (\ref{uspinisol}) and (\ref{Nspinav1}), and is given
by
\begin{eqnarray}
  \label{distantave}
 \langle {\cal U}_n(l)\rangle&=&l\left(1+\frac{\Delta}{2E_C}\right)
 +\left[(-1)^{n+l}-(-1)^n\right]\nonumber\\ 
 &\times&\frac{\Delta}{E_C}\left[-\frac{1}{2}+\frac{\ln 2}{2}\frac{T}{\Delta}
   -\sum_{\alpha=L,R}\gamma''_\alpha\right]\;.
\end{eqnarray}
Like in the case of neighboring peaks, which we considered in
Sec.~\ref{sec:posit}, the fluctuations of the peak spacings have the 
independent contributions $\delta{\cal U}'$, $\delta{\cal U}''$ and ${\cal
  U}^{(0)}$ coming from the randomness of the wave functions and
level spacings respectively.  Using Eqs.~(\ref{kvrhbi}), (\ref{UT})
and (\ref{deltaNHaldane}), we obtain for $l\ll E_C/\Delta$
\begin{eqnarray}
\label{distantfluct}
\langle [\delta{\cal U}'_n(l)+\delta{\cal U}''_n(l)]^2\rangle&=&\frac{4}{\beta}
\left(\frac{\Delta}{E_C}\right)^2\nonumber\\
&\times&\sum_{\alpha=L,R}
\left[4(l-1)(\gamma'_\alpha)^2+B_\alpha\right]
\;, 
\end{eqnarray}
where $B_\alpha$ depend only on whether the numbers $n$ and $n+l$ are even
or odd, and do not grow with $l$:
\begin{displaymath}
B_\alpha=\left\{
\begin{array}{ll}
(\gamma''_\alpha)^2\,&{\rm even}\ l \;,\\
(2\gamma'_\alpha+\gamma''_\alpha)^2-8(\gamma'_\alpha)^2\;,&
{\rm odd}\ l\geq 3,\ {\rm odd}\ n\;,\\
(\gamma''_\alpha)^2\;,&
l=1,\ {\rm odd}\ n\;,\\
(2\gamma'_\alpha-\gamma''_\alpha)^2\;,&{\rm odd}\ l,\ {\rm even}\ n\;.
\end{array}
\right.
\end{displaymath}
Thus the contribution due to the fluctuations of the wave functions,
$\delta{\cal U}'_n(l)+\delta{\cal U}''_n(l)$, is proportional to the
junction conductance, and its variance grows linearly with the
distance $l$ between the peaks.  This behavior is in qualitative
agreement with the experiments.\cite{MaurerEtal99} The fluctuations of
the other contribution, $\delta{\cal U}_n^{(0)}(l)$, do not depend on
the junction conductance, and are characterized by
\begin{equation}
  \label{distdelt}
\langle [\delta{\cal U}^{(0)}_n(l)]^2\rangle=
\left(\frac{\Delta}{E_C}\right)^2
F\left(\frac{l+(-1)^{n+l}-(-1)^n}{2}\right)\;.
\end{equation}
Function $F(k)$ determines the variance of the spacing between two
distant one-electron energy levels, and is given by the RMT;
$F(k)\sim\ln k$ at $k\gg 1$.
One can see that for sufficiently distant peaks and large junction
conductances, $\delta{\cal U}'(l)+\delta{\cal U}''(l)$ dominates
over $\delta{\cal U}^{(0)}(l)$, and the main term in the fluctuations of
the spacing comes from the fluctuations of the wave functions at the
points of dot-lead contacts.

At large $l$, the correlations between the sets of the levels that
determine the shifts of the $n$-th and $(n+l)$-th peaks become
negligible. If the Hamiltonian of the dot is not affected by adding of
new electrons, it occurs at $l\sim E_C/\Delta$, when these two sets of
levels, having characteristic width $E_C$ each, no more overlap. In
real dots, it can occur even at smaller $l$, because by adding
electrons one modifies the effective potential which confines
electrons to the dot. Eventually the modifications result in a loss of
correlations between the one-electron states in the original and new
potentials.  Then the positions of distant peaks are defined by two
Hamiltonians, both of the generic form (\ref{H})-(\ref{Ht}) with the
same $E_C$, but with different and statistically independent (random)
$t_{kp}$, $t_{kq}$ and $\varepsilon_k$. The resulting variance of the
peak spacing at large $l$ is twice the variance of a peak position
[Eq.~(\ref{deltaN})] and is given by
\begin{equation}
  \label{verydist}
\langle [\delta{\cal U}'_n(l)]^2\rangle=
\frac{1}{6\pi^2\beta}
\frac{\Delta}{E_C}
\sum_{\alpha=L,R}
g_\alpha^2\;.
\end{equation}
 The
contribution $\delta{\cal U}''_n(l)$ does not grow with $l$ and is
neglected here.

Thus the fluctuations of the interpeak spacing due to randomness of
the wave functions saturate at large distances between the peaks. The
corresponding fluctuation amplitude exceeds that for the neighboring
peaks by a parametrically large factor
$\sqrt{E_C/\Delta}/[\ln(E_C/\Delta)]^2$.  This saturation of the peak
spacing fluctuations with the increase of $l$ is in agreement with the
experimental results (see Fig.~2 in Ref.~\onlinecite{MaurerEtal99}).

As for the contribution of the level spacing fluctuations, ${\cal
  U}_n^{(0)}(l)$, its statistical properties for large $l$ depend
crucially on the nature of the scrambling of the one-electron states,
and are model-dependent; the discussion of this issue is beyond the
scope of the present paper. The experimental verification of
Eq.~(\ref{verydist}), however, is possible without knowledge of
$\langle [\delta{\cal U}_n^{(0)}(l)]^2\rangle$. Since $\delta{\cal
  U}_n^{(0)}(l)$ is independent on the junction conductance, the
difference between the values of $\langle [\delta{\cal
  U}_n(l)]^2\rangle$ measured at various junction conductances can be
directly compared to Eq.~(\ref{verydist}).

Equations (\ref{distantfluct}) and (\ref{verydist}) describe the
behavior of $\langle\delta{\cal U}^2_n(l)\rangle$ in the limiting
cases of small and large $l$ respectively. The crossover between these
two limits can be described analytically in the absence of scrambling of the
electron wave functions by adding new electrons to the dot.
Equations~(\ref{kvrhbi}), (\ref{UT}) and (\ref{deltaNHaldane}) yield 
\begin{equation}
  \label{newdistfluct}
\langle [\delta{\cal U}'_n(l)]^2\rangle=\frac{32}{\beta}
\left(\frac{\Delta}{E_C}\right)^2\nonumber\\
\sum_{\alpha=L,R}
\int_\Delta^{(1+l/2)\Delta}\frac{dD}{\Delta}\lambda_\alpha^2(D)\;,
\end{equation}
where $\lambda_\alpha(D)$ is given by Eq.~(\ref{lambdamufinal}).  At
small $l$, one can replace $\lambda_\alpha(D)$ in
Eq.~(\ref{newdistfluct}) with $\lambda_\alpha(\Delta)=\gamma'_\alpha$,
which yields the $l$-dependent part of Eq.~(\ref{distantfluct}).  For
$l\sim E_C/\Delta$, one can replace $\lambda_\alpha(D)$ with
$g_\alpha/(2\pi)^2$ within the most of the integration interval, so the
right-hand-side of Eq.~(\ref{newdistfluct}) matches
Eq.~(\ref{verydist}).


\section{Crossover between weak and strong tunneling}
\label{sec:crossover}

In the previous sections we considered the case of small enough
conductances of both dot-lead contacts [the condition
(\ref{RGCondition}) is satisfied].  In this weak tunneling regime, the
$n$-th conductance peak corresponds to the resonant tunneling via the
single, $n$-th level. The width of the peak is given by
\begin{equation}
  \label{weakwidth}
  {\cal W}_n\sim 
\frac{\max\left\{\Gamma_{nL}+\Gamma_{nR}, T\right\}}{E_C}\;, 
\end{equation}
where $\Gamma_{n\alpha}$ is the partial width of the $n$-th level with respect
to the electron tunneling to the $\alpha$-th lead. If the coupling of
the dot to the leads is weak enough so the condition
(\ref{ptCondition}) is satisfied, the partial level widths are given by the
conventional expression
\begin{equation}
  \label{Gamma1}
  \Gamma_{n\alpha}=g_\alpha
\frac{|\psi_n({\bf r}_\alpha)|^2}{\langle|\psi_n|^2\rangle}
\frac{\Delta}{2\pi}\;.
\end{equation}
At larger values of $\sqrt{g_\alpha}\ln(E_C/\Delta)$, the
level widths are renormalized (see Sec.~\ref{sec:RG}):
\begin{equation}
  \label{Gamma2}
  \Gamma_{n\alpha}=[2\pi\mu_\alpha(\Delta)]^2
\frac{|\psi_n({\bf r}_\alpha)|^2}{\langle|\psi_n|^2\rangle}
\frac{\Delta}{2\pi}\;.
\end{equation}
It implies the renormalization of the junction conductance:
\begin{equation}
\label{renormcond}
  g_\alpha(D)=[2\pi\mu_\alpha(D)]^2\;,
\end{equation}
where $\mu_\alpha(D)$ given by Eq.~(\ref{lambdamufinal}).

The position of a conductance peak is mainly determined by the
position of the corresponding charge degeneracy point. However, there
is another random contribution to the shift of the peak position,
which stems from the randomness of the phases of the transmission
amplitudes via different one-electron levels in the dot. Even when the
dot is in the resonance, and the dominating contribution to the
tunneling amplitude comes from a single resonant level, there is also
a contribution from the cotunneling\cite{AverinNazarov90} through the
other levels, which yields the following estimate for the tunneling
amplitude:
\begin{eqnarray}
{\cal A}_n(\delta{\cal N})&=&
e^{i\eta_n}
\sqrt{\frac{\Gamma_{nL}\Gamma_{nR}}
           {[(\Gamma_{nL}+\Gamma_{nR})/2]^2+(E_C\delta{\cal N})^2}
     }  \nonumber\\
&+&\sum_{k\neq n}
   e^{i\eta_k}
   \frac{\sqrt{\Gamma_{kL}\Gamma_{kR}}}
        {\varepsilon_k-\varepsilon_n+E_C\delta{\cal N}}\;,
\label{fullamp}
\end{eqnarray}
where $\eta_k$ is the random phase of the amplitude of tunneling
through the $k$-th level in the dot, and $\delta{\cal N}$ is the
deviation from the closest charge degeneracy point.  The position of
the maximum of $|{\cal A}_n(\delta{\cal N})|$ given by
Eq.~(\ref{fullamp}), $\delta {\cal N}=\delta {\cal N}_n$, may not
coincide with the point $\delta {\cal N}=0$, {\em i.e.}  the
cotunneling contribution adds to the random shift of a peak.
Equation~(\ref{fullamp}), together with Eq.~(\ref{Gamma2}), yields the
following estimate for the corresponding contribution to the peak
spacing fluctuations:
\begin{equation}
  \label{shiftetm}
  \delta {\cal U}^{\rm (cotunn)}_n\sim
  \frac{\Delta}{E_C}\left[\mu_L^2(\Delta)+\mu_R^2(\Delta)\right]^3\;.
\end{equation}

This contribution is much smaller than the one coming from the
fluctuations of the charge degeneracy points [Eqs.~(\ref{Uspinprim}),
(\ref{Uspinprim})], provided condition (\ref{RGCondition}) is
satisfied. This allowed us to neglect the contribution
(\ref{shiftetm}) yet so far.

Now we would like to extend our consideration beyond the limits set by
Eq.~(\ref{RGCondition}). To demonstrate the new physical features
emerging at stronger coupling of the dot to the leads, it is
sufficient to limit ourselves to the case when the conductance of one
channel (say, the right one) is small, and satisfies
Eq.~(\ref{ptCondition}), as the other one ($g_L$) varies between $0$
and $1$. The consideration of the whole 2-dimensional parameter
space $0<g_L<1$, $0<g_R<1$ would not add anything new to our
understanding of the nature of the fluctuations of the conductance
peak positions and will be omitted.

To begin with, we consider the conductance of the dot in the
conditions opposite to the limit of weak coupling,
\begin{equation}
g_R\ll 1\;,\;\; 1-g_L\ll 1\;.
\label{stronggg}
\end{equation}
This case was studied in details in Ref.~\onlinecite{AleinerGlazman98}
and we first recall some findings of this work. Since one of the
contacts is almost open, the charge of the dot is not quantized
anymore.  The Coulomb blockade is almost lifted and the notion of the
charge degeneracy points is useless. Instead of sharp peaks, the
conductance as a function of gate voltage ${\cal N}$ exhibits
oscillations with the period of unity.  The oscillations occur around
the value 
\begin{equation}
  \label{charlar}
 \langle G\rangle=G_0 g_R\frac{\Delta}{E_C} \;,
\end{equation}
the characteristic amplitude of the oscillations is also
$G_0g_R(\Delta/E_C)$. The phase of oscillations is a random quantity,
varying with ${\cal N}$ slowly (at scales larger that 1).  The
electron transport through the dot is mediated not by a single
resonant level, as in the case of weak coupling, but by a set of
levels, whose energies lie in the strip of the width $E_C$ near the
Fermi energy. As a result, the value of conductance at given gate
voltage ${\cal N}$ is determined by the properties of all levels in
this set. Since there are no well-pronounced peaks, it is more natural
to describe the statistical properties of conductance not in terms of
the heights and positions of the conductance maxima, but rather in
terms of the correlation functions $\langle G({\cal N}_1) G({\cal
  N}_2)\rangle$.

However, we still would like to study the statistical behavior of the
conductance maxima, following the experiments,\cite{MaurerEtal99} and
in order to make a connection to the case of weak conductance.  The
quantitative consideration is hampered by the fact that, in the absence of
sharp peaks, the exact positions of the conductance maxima are not
associated with special points of any physical quantity. Therefore we
will limit ourselves to a qualitative consideration of the
fluctuations. It was mentioned above, that the Coulomb blockade is
lifted when $g_L\approx 1$.  Under these conditions, the conductance
maxima occur at the points where the amplitudes of tunneling via
$E_C/\Delta$ levels in the dot add up most favorably [Cf.
Eqs.~(\ref{fullamp}),~(\ref{shiftetm})]. The phase of transmission via
each level, $\eta_k$, depends on the gate voltage, so the function
$G({\cal N})$ is not constant. It is (almost) periodic, because the
functions $\eta_k({\cal N})$ have period of unity; $G({\cal N})$ is
not exactly periodic because when the gate voltage ${\cal N}$ is
shifted by unity, one level is replaced in the set which mediates
conductance.  Therefore, the conductance maxima are not equally
spaced, but their positions fluctuate with respect to each other.
When ${\cal N}$ is varied by $E_C/\Delta$, all the ``conducting''
levels are replaced, so $G({\cal N})$ is not correlated with $G({\cal
  N}+E_C/\Delta)$.  In particular, the positions of the maxima
separated by $E_C/\Delta$ other maxima fluctuate independently within
range $\delta {\cal N}_n\sim 1$ each.  Therefore, the characteristic
amplitude of fluctuations of the spacing between nearest maxima equals
\begin{equation}
  \label{sdfg}
\delta{\cal U}_n\sim \sqrt{\frac{\Delta}{E_C}}
\end{equation}
for the conditions defined in Eq.~(\ref{stronggg}).

Now we can combine the knowledge acquired in the consideration of the
case of weak [Eq.~(\ref{RGCondition})] and strong
[Eq.~(\ref{stronggg})] tunneling through the left contact, to deal with
the case of the intermediate conductance
\begin{equation}
  \label{interG}
  \left[\frac{\pi}{\ln(E_C/\Delta)}\right]^2<g_L< 1\;.
\end{equation}
In considerations of Sec.~\ref{sec:posit} we used the renormalization
group transformation of the Hamiltonian only in the very vicinity of
the conductance peak. In principle, this transformation can be
performed at any value of $\delta{\cal N}$, where $\delta{\cal N}$ is
the distance to the closest charge degeneracy point of Hamiltonian
(\ref{H}).  If
\begin{equation}
\label{cond1}
|\delta{\cal N}|>\exp\left(-\pi/\sqrt{g_L}\right)  \;,
\end{equation}
then the renormalization must be stopped when the reduced band width
$D$ reaches $E_C|\delta{\cal N}|$.  Indeed, if $D<E_C|\delta{\cal
  N}|$, then one of the two dot's charge states is left beyond the
band limits; only one charge state is then possible.  One can see that
the renormalization defined by Eq.~(\ref{Tscaling}) does not bring the
system out of the conditions of weak coupling, because
$g_L(D=E_C|\delta{\cal N}|)$ is still much less then unity.  The
conductance of the dot can then be estimated from $\hat{\sf
  H}(D=E_C|\delta{\cal N}|)$ with the help of the formula for elastic
cotunneling. Under the condition (\ref{cond1}), the conductance is
\begin{equation}
    \label{intervalley}
G\sim G_0
g_R g_L(D)
\frac{\Delta}{D}\;,
\quad\mbox{with}\quad D=E_C|\delta{\cal N}|\;.
\end{equation}
The renormalization of the conductance of the right
contact, $g_R$, can be neglected, since it satisfies Eq.~(\ref{ptCondition}).

If, on the contrary, 
\begin{equation}
\label{cond2}
|\delta{\cal N}|<\exp\left(-\pi/\sqrt{g_L}\right)
\;,
\end{equation}
then the renormalization is stopped when $D$ reaches
$E_C\exp(-\pi/\sqrt{g_L})$. At this point, the renormalized
conductance of the left contact, given by Eq.~(\ref{renormcond}),
becomes of the order of unity. Then the scaling law~(\ref{Tscaling}),
based on the perturbation theory in the small parameter $g_L(D)$,
is not applicable. The nature of conductance through the dot with
such parameters (one contact is almost open, the other one is almost
closed) was discussed earlier in this section. The characteristic
conductance for the conditions of Eq.~(\ref{cond1}) is given by the
formula similar to Eq.~(\ref{charlar}). In this formula, $E_C$, which
in Eq.~(\ref{charlar}) plays the role of the width of the band of
electrons affected by the charging, must be replaced with
$D=E_C\exp(-\pi/\sqrt{g_L})$:
\begin{eqnarray}
  G\sim 
G_0 g_R
  \frac{\Delta}{E_C}\exp\left(\frac{\pi}{\sqrt{g_L}}\right)\;.
  \label{Gintpeak}
\end{eqnarray}
This value is parametrically larger than the one given by
Eq.~(\ref{intervalley}). Thus the characteristic width of a
conductance peak in the conditions of Eq.~(\ref{interG})
is 
\begin{equation}
  \label{intwidth}
  {\cal W}_n\sim \exp\left(-\frac{\pi}{\sqrt{g_L}}\right)\;.
\end{equation}
 The conductance maxima are not tied to
the charge degeneracy points. Instead, their positions determined by
the interference of the amplitudes of transmission through
$(E_C/\Delta)\exp(-\pi/\sqrt{g_L})$ levels, that belong to the
reduced band of width $E_C\exp(-\pi/\sqrt{g_L})$.  Therefore, the
conductance may have a maximum, roughly speaking, at any point lying
in the interval where the renormalized conductance of the left contact
is close to unity, {\em i.e.}, within the region defined by
$|\delta{\cal N}|<\exp(-\pi/\sqrt{g_L})$. The positions of the
maxima corresponding to two neighboring charge degeneracy points are
correlated, because they are defined by the same set of levels except
for one. This situation is similar to that in the case when one of the
contacts {\em initially} had conductance close to unity. The
difference is that now the set of the levels which determines the
position of a maximum is smaller. Employing the same ideas we used
for the dot with one contact initially almost open, we obtain that the
characteristic amplitude of fluctuations of the maxima spacings  is
\begin{equation}
  \label{dfsds}
\delta{\cal U}_n\sim
\sqrt{\frac{\Delta}{E_C}}
\exp\left(-\frac{\pi}{2\sqrt{g_L}}\right)
\;,
\end{equation}
for the conditions defined by Eq.~(\ref{interG}).  This expression
matches Eqs.~(\ref{varspin}) and (\ref{sdfg}) at $g_L\sim 
[\pi/\ln(E_C/\Delta)]^2$ and $g_L\sim 1$, respectively.

Equation~(\ref{dfsds}) gives an estimate for the fluctuations of
spacing between two neighboring maxima. Now we extend this estimate to
distant maxima.  As we discussed before in this Section, the random
positions of the conductance maxima are determined by the sets of
$\sim (E_C/\Delta)\exp(-\pi/\sqrt{g_L})$ levels. For the $n$-th and
$(n+l)$-th maxima, the levels in these two sets are the same except
for $l$ of them. These $l$ levels in each of the two sets produce
uncorrelated contributions to the positions of the two maxima. This
contribution grows with $l$, since the contribution of each particular
level is independent of the others:
\begin{eqnarray}
  \label{sdfg-distant}
\delta{\cal U}_n(l)&\sim&\sqrt{l}\:
\sqrt{\frac{\Delta}{E_C}}
\exp\left(-\frac{\pi}{2\sqrt{g_L}}\right)\;,\\
&&\quad \mbox{for}\quad l
<\frac{E_C}{\Delta}\exp\left(-\frac{\pi}{\sqrt{g_L}}\right)\;.
\nonumber
\end{eqnarray}
At $l>(E_C/\Delta)\exp(-\pi/\sqrt{g_L})\;$ there are no
common levels in the two sets, and the fluctuations of the positions
of the $n$-th and $(n+l)$-th peaks are independent. The variance of
spacing between them saturates at large $l$:
\begin{eqnarray}
  \label{sdfg-verydistant}
  \delta{\cal U}_n(l)&\sim&
  \exp\left(-\frac{\pi}{\sqrt{g_L}}\right)\;,\\
&&\quad \mbox{for}\quad
  l>\frac{E_C}{\Delta}\exp\left(-\frac{\pi}{\sqrt{g_L}}\right)
\;.\nonumber
\end{eqnarray}

The levels that are eliminated by the reduction of the band down to
$E_C\exp(-\pi/\sqrt{g_L})$ in the course of renormalization also
make a contribution to $\delta{\cal U}_n(l)$.  However, it is
important only in the narrow region of crossover between the regimes
of weak [Eqs.~(\ref{newdistfluct}), (\ref{verydist})] and intermediate
[Eqs.~(\ref{sdfg-distant}), (\ref{sdfg-verydistant})] conductance,
where the number of levels in the reduced band is of the order of
unity.  Otherwise this contribution is small, as compared to the one
given by Eq.~(\ref{sdfg-distant}), and can be neglected.


\section{Conclusion}

We have studied the statistical properties of the spacings between the
conductance peaks in a partially open Coulomb blockaded quantum dot.
We found that the fluctuations of the electron wave functions in the
dot contribute to the fluctuations of the peak spacings; this
contribution grows with the conductance of the contacts, see
Table~\ref{table}.

For relatively weak coupling of the dot to the leads,
$g_{L,R}<[\pi/\ln(E_C/\Delta)]^2$, the peak conductance is mediated
by a single electron level, and the peaks are sharp. The variance of
the peak spacing fluctuations for the neighboring peaks is given by
Eqs.~(\ref{varspin}); it reaches $\Delta^2$ when
$g_{L,R}\sim[\pi/\ln(E_C/\Delta)]^2$.

At larger values of $g_{L,R}$, the conductance peaks are broad, and
many levels are involved in the transport through the dot.  Thus the
properties of conductance peaks (more precisely, maxima) are
determined by the wave functions of many electron states. The variance
of the spacing between the maxima continues to grow with the
conductance of contacts, and is given by Eq.~(\ref{dfsds}).

The fluctuations of the spacing between two conductance peaks grow
with the distance between them, see Eqs.~(\ref{distantfluct}),
(\ref{sdfg-distant}). At large distances between two peaks, the variance
of the spacing between them saturates at the value given by
Eq.~(\ref{verydist})~(\ref{sdfg-verydistant}).

The dependence of the peak position on the one-electron wave functions
in the dot leads to the correlations between the fluctuations in the
peaks' spacings and heights: the spacings between the higher peaks
are, on average, smaller. The parameter characterizing this
correlation is given by Eq.~(\ref{corr}).
\end{multicols}

\widetext
\begin{table}
  \caption{The characteristic values of the peak width and the
    amplitude of fluctuations of the spacing between the neighboring
    peaks in units of $E_C$ for $g_R\ll 1$, $g_R<g_L< 1$, as 
    given by Eqs.~\protect~(\ref{varspin}), (\protect\ref{Gamma2}),
    (\protect\ref{intwidth}), (\protect\ref{dfsds}). The numeric
    factors are omitted.  }
  \begin{tabular}{ccc}
conductance of the contacts
& peak width & $\displaystyle\sqrt{\langle\delta{\cal U}^2_n\rangle}$
\rule[-0.1in]{0in}{0.2in}
\\
\hline
$g_R
<g_L
<\left[\displaystyle\frac{\pi}{\ln(E_C/\Delta)}\right]^2$&
$\displaystyle\frac{\Delta}{E_C}g_L$&
$\displaystyle\frac{\Delta}{E_C}
\sqrt{g_L}
\tan\left(\displaystyle\frac{1}{\pi}\sqrt{g_L}
\ln\displaystyle\frac{E_C}{\Delta}\right)$
\rule[-0.1in]{0in}{0.37in}
\\
$\left[\displaystyle\frac{\pi}{\ln(E_C/\Delta)}\right]^2
<g_L<1\;,\quad g_R<g_L$&
$\exp\left(-\displaystyle\frac{\pi}{\sqrt{g_L}}\right)$&
$\sqrt{\displaystyle\frac{\Delta}{E_C}}
\exp\left(-\displaystyle\frac{\pi}{2\sqrt{g_L}}\right)$
\rule[-0.17in]{0in}{0.41in}
  \end{tabular}

  \label{table}
\end{table}

\begin{multicols}{2}

\acknowledgements

This work has been supported by the NSF Grant No. DMR-9731756. The
discussions with C.M. Marcus are gratefully acknowledged.

\end{multicols}

\begin{references} 
\bibitem{Kastner92} M.A. Kastner, Rev. Mod. Phys. {\bf 64}, 849 (1992).
\bibitem{MehtaBook} M.L. Mehta, {\em Random Matrices}, 2nd
  ed. (Academic Press, London, 1991).
\bibitem{JalabertEtal92} R.A. Jalabert, A.D. Stone, and Y. Alhassid,
  Phys. Rev. Lett. {\bf 68}, 3468 (1992).
\bibitem{ChangEtal96}A.M. Chang {\em et al.}, Phys. Rev. Lett.  {\bf
    76}, 1695 (1996); J.A. Folk {\em et al.}, {\em ibid.}  {\bf
    76}, 1699 (1996); S.M. Cronenwett {\em et al.}, {\em ibid.}
  {\bf 79}, 2312 (1997).  
\bibitem{SivanEtal96} U. Sivan {\em et al.}, Phys. Rev. Lett. {\bf
  77}, 1123 (1996).
\bibitem{SimmelEtal97}F. Simmel, T. Heinzel, and D.A. Wharam,
  Europhys. Lett. {\bf 38}, 123 (1997); F. Simmel {\em et al.},
  Phys. Rev. B {\bf 59}, R10441 (1999).
\bibitem{PatelEtal98} S.R. Patel {\em et al.}, Phys. Rev. Lett. {\bf
  80}, 4522 (1998).
\bibitem{WalkerEtal99} See P.N. Walker, Y. Gefen, G. Montambaux,
  Phys. Rev. Lett. {\bf 82}, 5329 (1999) and references therein.
\bibitem{VallejosEtal98} R.O. Vallejos, C.H. Lewenkopf, E.R. Mucciolo,
Phys. Rev. Lett. {\bf 81}, 677 (1998).
\bibitem{MaurerEtal99} S.M. Maurer {\em et al.}, Phys. Rev. Lett.
 {\bf 83}, 1403 (1999).
\bibitem{Anderson70} P.W. Anderson, J. Phys. C {\bf 11}, 5015 (1978).
\bibitem{GlazmanMatveev89} L.I. Glazman and K.A. Matveev, Sov. Phys.
  JETP {\bf 71}, 1031 (1990); K.A. Matveev {\em ibid.}  {\bf72}, 892
  (1991).  
\bibitem{Beenakker91} C.W.J. Beenakker, Phys. Rev. B {\bf 44}, 1646 (1991).
\bibitem{Haldane78} These quantities are analogous to the corrections
  to the energy levels of an Anderson impurity, and were calculated in:
 F.D.M. Haldane, Phys. Rev. Lett. {\bf 40}, 416 (1978).
\bibitem{PatelEtal98-2} S.R. Patel {\em et al.}, Phys. Rev. Lett.
{\bf 81}, 5900 (1998); D.R. Stewart {\em et al.}, Science {\bf 278},
1784 (1997).
\bibitem{AverinNazarov90} D.V. Averin, Yu.V. Nazarov, Phys. Rev. Lett.
  {\bf 65}, 2446 (1990).
\bibitem{AleinerGlazman98} I.L. Aleiner, L.I. Glazman, Phys. Rev. B
  {\bf 57}, 9608 (1998).  
\end{references}
\end{document}